\documentstyle[aps,prb,epsfig,floats]{revtex}
\begin{document}

\twocolumn[\hsize\textwidth\columnwidth\hsize\csname@twocolumnfalse\endcsname

\title{Lattice dielectric response of 
$\mbox{CdCu}_{3}\mbox{Ti}_{4}\mbox{O}_{12}$ and of
$\mbox{CaCu}_{3}\mbox{Ti}_{4}\mbox{O}_{12}$ from first principles}
\author{Lixin He, J. B. Neaton, David Vanderbilt, and Morrel H. Cohen}
\address{Department of Physics and Astronomy, Rutgers University,\\
Piscataway, New Jersey 08855-0849}
\date{September 9, 2002}

\draft
\maketitle

\begin{abstract}

Structural, vibrational, and lattice dielectric properties of
CdCu$_3$Ti$_4$O$_{12}$ are studied using density-functional theory
within the local spin-density approximation, and the results are
compared with those computed previously for CaCu$_3$Ti$_4$O$_{12}$.
Replacing Ca with Cd is found to leave many calculated quantities largely
unaltered, although significant differences do emerge in
zone-center optical phonon frequencies and mode effective charges.
The computed phonon frequencies of CdCu$_3$Ti$_4$O$_{12}$ are found
to be in excellent agreement with experiment, and the
computed lattice contribution to the intrinsic static dielectric constant
($\sim$60) also agrees exceptionally well with a recent optical
absorption experiment. These results provide further support for
a picture in which the lattice dielectric response is essentially
conventional, suggesting an extrinsic origin for the anomalous
low-frequency dielectric response recently observed in both materials.
\end{abstract}


\narrowtext

]

\narrowtext

The complex perovskite compound CaCu$_3$Ti$_4$O$_{12}$ (CCTO) has
recently attracted considerable attention in view of its anomalously
large dielectric response.  Low-frequency studies uncovered dielectric
constants near 16,000 for ceramics\cite{sub,ram} and 80,000 for
single-crystal samples\cite{homes} that are {\it nearly constant} 
over a wide temperature range ($\sim$100-600$\,$K).  
Moreover, above a critical frequency that ranges
between $\sim$10$\,$Hz and 1$\,$MHz depending on temperature, the
dielectric constant drops to a rather ordinary value of $\sim$100,
typical of conventional perovskite insulators\cite{lg} where
contributions from zone-center infrared-active phonons can easily
result in values of this magnitude. Detailed measurements of the dynamic 
susceptibility indicated that the low-temperature 
crossover is characteristic of an activated Debye-type relaxation 
process.\cite{homes} 

In a previous first-principles study of CCTO,\cite{lixin} we
calculated the lattice contribution to the dielectric constant,
and found it to be in order-of-magnitude agreement with the
measured far-infrared value\cite{homes} and drastically lower
than the static value.  Thus, while the dielectric response
in the infrared range is essentially well understood, the origin of
the enormous low-frequency response and its dynamic behavior
remain perplexing. An extrinsic mechanism, i.e., one associated
with defects, domain boundaries, or other crystalline deficiencies,
was initially proposed by Subramanian {\it et al.}\cite{sub} These
authors reported twinning in CCTO samples and suggested that
internal capacitive barrier layers, formed by domain boundaries,
might explain the unusual dielectric response.  A similar
explanation has been proposed by another group.\cite{sinclair}
We discussed some of these issues in Ref.\onlinecite{lixin} and, in
the absence of direct evidence for intrinsic lattice or electronic
origins, we concluded that extrinsic effects were the likely source
of the unusual low-frequency behavior. A subsequent report by
Lunkenheimer {\it et al}.\cite{lunken} discussed a scenario in
which barrier layers associated with poor metallic contacts are
responsible for the Debye relaxation.  Further discussion of
extrinsic mechanisms will be given elsewhere.\cite{morrel}

While the enormous low-frequency response is the primary puzzle
in this material, two other aspects of the response 
at higher frequencies are also currently enigmatic.
First, the oscillator strength of one low frequency IR-active 
mode has been observed to {\it
increase} strongly with decreasing temperature.  And second, one
IR-active mode, at somewhat higher frequency, is not observed at
all, although it is expected by symmetry and has
been predicted to possess a non-vanishing oscillator
strength.\cite{lixin,homes2}  It is not clear whether there
could be any connection between these IR anomalies and
the enormous low-frequency response, but in any case,
further experimental and theoretical scrutiny of the
infrared phonon spectrum would clearly be of significant physical
interest.\cite{homes}

In this report, we extend our previous work on CCTO by carrying out
a parallel study on the closely related material
CdCu$_3$Ti$_4$O$_{12}$ (CdCTO).\cite{sub,homes2}  A recent
experimental study of ceramic CdCTO samples\cite{homes2} has
revealed that CdCTO also has a temperature-independent anomalous
response over a wide range and shows a similar Debye crossover
between low- and high-frequency regimes.  While the magnitude of
this response ($\epsilon\sim$400 for ceramic samples) 
is considerable, it is nonetheless significantly
lower than in CCTO, which is somewhat surprising since Cd and Ca
are chemically so similar.  (Anomalies also appear in the
low-frequency IR-active mode oscillator strengths of CdCTO.)
After calculating ground-state and dielectric properties
of CdCTO, we compare with recent experiments and with the corresponding
quantities for CCTO. The results obtained for both materials are
similar, typical of ordinary perovskite oxides, and in
excellent agreement with recent optical
experiments.\cite{homes,homes2} Our results indicate no significant
fundamental differences in the intrinsic properties of these two
materials and thus provide support for an extrinsic origin of their
substantial low-frequency responses.

Our first-principles study of CdCTO employs density-functional
theory within the local spin-density approximation (LSDA), as
implemented within the Vienna {\it ab-initio} Simulations Package
(VASP).\cite{kresse1,kresse2} VASP utilizes a plane-wave basis and
Vanderbilt ultrasoft pseudopotentials.\cite{vanderbilt} All
pseudopotentials include non-linear core corrections.\cite{froyen}
For Ca and Ti, these potentials treat the highest occupied $p$
shell electrons explicitly as valence; for Cu, we also consider
electrons in the $3d$ and $4s$ shells self-consistently; and for
Cd, the $4d$ and $5s$ shells are included explicitly.  
The ions are relaxed towards equilibrium until
the Hellmann-Feynman forces are less than 10$^{-3}$\,eV/{\AA}.
Brillouin-zone integrations are performed with a Gaussian
broadening of 0.1\,eV during all relaxations.  A 37\,Ry
plane-wave cut-off and a 2$\times$2$\times$2 Monkhorst-Pack {\bf k}-point
mesh (equivalent to a 4$\times$4$\times$4 mesh for a single 5-atom
perovskite unit cell) results in good convergence of all
properties reported here.

%---------------------------------------------------------------------

\begin{table}
\caption{Comparison of calculated and measured structural
parameters of CCTO and CdCTO. Both have the space group $Im3$
(point group $T_h$); the Wyckoff positions are Ca/Cd($0,0,0$),
Cu($1/2,0,0$), Ti($1/4,1/4,1/4$), O($x,y,0$). 
See Ref.~\protect\onlinecite{volume} for further
details. \label{tab:structure}}
\vskip 0.1cm
\begin{tabular}{l|cdd}
                      & Structural parameter & LSDA       & Exp.\\
\hline
         &$a$ (\AA)  &   7.324    & 7.384\tablenotemark[1]   \\
CaCu$_3$Ti$_4$O$_{12}$  & O($x$)         &   0.303    & 0.303   \\
                      &O($y$)         &   0.175    & 0.179   \\
\hline
         &$a$ (\AA)  &   7.324    & 7.384\tablenotemark[2]  \\
CdCu$_3$Ti$_4$O$_{12}$  & O($x$)         &   0.306    &   \\
                      &O($y$)         &   0.176    &    

\end{tabular}
\tablenotetext[1] {35\,K ($a$=7.391\,\AA\, at 298\,K).}
\tablenotetext[2] {298\,K.}
\end{table}

%---------------------------------------------------------------------

Neutron-diffraction measurements\cite{sub} suggest that, as with
CCTO, CdCTO crystallizes in a lattice with a 20-atom body-centered
cubic primitive cell\cite{bochu} having space and point groups
$Im3$ and $T_h$, respectively. Since the $Im3$ space group is
centrosymmetric, the spontaneous polarization necessarily vanishes
by symmetry.  To accommodate the observed antiferromagnetic
(AFM) spin arrangement, all calculations are performed using a
doubled 40-atom simple-cubic unit cell containing eight perovskite
5-atom units, of which 3/4 have Cu on the A site and 1/4 have
Ca or Cd on the A site. (Each Cu--Cu nearest-neighbor pair has
antiparallel spins.)

Structural parameters resulting from the relaxation appear in
Table I.  Replacing Ca with Cd leaves the lattice constant virtually
unchanged,\cite{volume} and the internal parameters also remain
essentially the same.  Only minor structural changes would be
expected, given the similar nominal valence ($+$2) and ionic radii
of Ca and Cd (1.48 and 1.45\,\AA, respectively\cite{shannon}).
Unfortunately we are unaware of any experimental refinements of the
internal structural parameters of CdCTO, and thus direct comparison
of our computed oxygen positions with experiment is not possible.  
As for the lattice parameters, however, our
LSDA results are consistent with the near-negligible
difference ($<$0.1\%) between the lattice constants measured for
CCTO and CdCTO (7.391 and 7.384\,\AA\ at 298\,K, respectively).
(The CCT0 lattice constant at 35~K is smaller and, coincidentally,
7.384\,\AA; a low-temperature lattice constant for CdCTO has yet
to be measured.)  
Lattice constants computed within the generalized
gradient approximation\cite{pw91} (GGA) are calculated to be about 1.2\%
larger than experiment for both CCTO and CdCTO.
All results presented below were computed within the LSDA without
gradient corrections.

The computed electronic structure (at $T=0$) is quite similar for
both materials; replacing Ca with Cd evidently has little
effect on electronic and magnetic properties.  CdCTO is found to
possess an antiferromagnetic (AFM) insulating ground state, and its
single-particle density of states (DOS)
closely corresponds to that of CCTO near the
band edges (see Figs.~2 and 3 in Ref.~\onlinecite{lixin}). 
Although the Cd ion has a nominal valence of
$+$2, its electronic configuration includes a filled $4d$ shell,
which results in a well-localized set of $d$ bands located about
8~eV below the valence band maximum.  As with CCTO, the gap and
magnetic moment of CdCTO originate from a splitting of hybridized
Cu($3d$)-O($2p$) $\sigma$-antibonding states.  The magnetic moment
of each CuO$_4$ plaquette, estimated from spin densities, is
$\sim$0.84\,$\mu_B$, and the (indirect) band gap is computed to be
about 0.19\,eV. To our knowledge, experimental values for the optical
gap remain unavailable.  However, the observed gap will undoubtedly be
strongly underestimated by our calculations, as was found for CCTO,
owing to the well-known limitations of the LSDA.

The lattice contribution to the dielectric constant that we computed
for CCTO ($\sim$40) was roughly a factor of two smaller than the value
measured by far-infrared spectroscopy.\cite{homes}
We now perform a similar calculation for
CdCTO and carefully examine any differences between the materials.

The static {\it lattice} dielectric response
can be approximated (neglecting anharmonicity) 
as the zero-frequency response
of a system of classical Lorentz oscillators, i.e.,
\begin{equation}
\epsilon_{\rm ph} = {\Omega_0}^2 \sum_\lambda
{{{Z^*_\lambda}^2}\over{{\omega_\lambda}^2}} \;,
\end{equation}
where  $\omega_\lambda$ and $Z^*_{\lambda}$ are
respectively the IR-active mode frequencies and mode dynamical
charges, and ${\Omega_0}^2= 4\pi e^2/m_0 V$ is a characteristic
frequency having the interpretation of a plasma frequency of a gas
of objects of mass $m_0$=1\,amu, charge $e$, and density $V^{-1}$
($V$ is the 20-atom primitive cell volume).

The lattice dielectric constant $\epsilon_{\rm ph}$ is obtained
by first calculating the frequency and mode effective charges of
each zone-center IR-active phonon, and then inserting these
quantities into Eq.~(1).  Since CdCTO and CCTO possess the same
structure, their zone-center phonons are computed and analyzed as
in our previous study.\cite{lixin} As discussed there, of the six
irreducible representations allowed by the $T_h$ point group, only
the $\mbox{T}_u$ modes display IR activity.  After
obtaining symmetry-adapted modes of $\mbox{T}_u$
symmetry, IR-active phonon frequencies and corresponding
eigenvectors are calculated using the frozen-phonon method, as
described in detail elsewhere.\cite{lixin}  The results
appear in Table II, where we also compare with corresponding
theoretical results for CCTO\cite{phonon} and with recent experimental
data.  As was the case for CCTO, all CdCTO IR-active modes
are found to be stable ($\omega^2>0$) and their computed
frequencies agree very well with experiment.\cite{homes2}

%---------------------------------------------------------------------
\begin{table}
\caption{Comparison of calculated mode frequencies $\omega_\lambda$,
effective charges $Z_{\lambda}^*$, and oscillator strengths
$S_{\lambda}=\Omega_0^2\,Z_{\lambda}^{*2}/\omega_{\lambda}^2$
of IR-active T$_u$ modes with the experimental values (at $T=295$~K) of
Homes {\it et al.}~(Refs.~\protect\onlinecite{homes,homes2}).
See Ref.~\protect\onlinecite{phonon} for further details.}
\label{tab:IR}
\vskip 0.1cm
\begin{tabular}{dddddddd}
\multicolumn{4}{c}{CaCu$_3$Ti$_4$O$_{12}$} &
\multicolumn{4}{c}{CdCu$_3$Ti$_4$O$_{12}$} \\
\multicolumn{2}{c}{$\omega$ (cm$^{-1}$)} &
\multicolumn{2}{c}{$S_\lambda$} &
\multicolumn{2}{c}{$\omega$ (cm$^{-1}$)} &
\multicolumn{2}{c}{$S_\lambda$} \\
LSDA & \multicolumn{1}{c}{Exper.} &
LSDA & \multicolumn{1}{c}{Exper.} &
LSDA & \multicolumn{1}{c}{Exper.} &
LSDA & \multicolumn{1}{c}{Exper.} \\
\hline
125 & 122.3 &  4.6 & 14.3 &72	   & 74     & 15.1   &  22.8\\ 
135 & 140.8 &  9.3 & 15.9 &125	   & 121    &  7.3   &  11.3\\  
158 & 160.8 &  7.5 & 6.92 &148     & 155    &  0.4   &   7.9\\ 
199 & 198.9 &  2.7 & 5.25 &170     & 167    &  11.8  &   4.3\\
261 & 253.9 & 12.5 & 13.8 &238     & 239    &  12.6  &   7.9\\  
310 & 307.6 &  1.0 & 0.68 &303     & 295    &  2.4   &   2.8\\   
385 & 382.1 &  0.5 & 1.96 &385     & 385    &  0.7   &   1.6\\ 
416 & 421.0 &  5.5 & 1.72 &405     & 422    &  2.4   &   1.4\\ 
471 &       &  1.5 &      &461     & 469    &  4.2   &   0.3\\ 
494 & 504.2 &  0.6 & 0.78 &499     & 494    &  0.4   &   1.1\\
547 & 552.4 &  0.4 & 0.62 & 545    & 550    &  0.3   &   0.7 
\end{tabular}
\end{table}
%---------------------------------------------------------------------

Given that Ca and Cd are isoelectronic, and that CCTO and CdCTO
possess similar structural and electronic properties, we may expect
their phonon frequencies and eigenvectors to be similar as well,
especially for modes that do not involve Ca or Cd.
To investigate this and related issues, we find it useful to
sort the zone-center mode frequencies from lowest to
highest in frequency, and then project each CCTO eigenvector onto
the corresponding one of CdCTO.
For all but the lowest four modes, the projections result in values
$>$~0.9, reflecting a close correspondence between the character of
their eigenmodes. However, the third- and fourth-lowest
eigenvectors (the 148 and 170\,cm$^{-1}$ modes of CdCTO and the 158
and 199\,cm$^{-1}$ modes of CCTO) are substantially mixed by
replacement of Ca with Cd, and possess considerably different
eigenvectors. Further, the frequencies of the lowest two modes reverse
order on going to the Cd from the Ca material:  the 72\,cm$^{-1}$
mode of CdCTO shares nearly the same eigenvector with the 141\,cm$^{-1}$
mode in CCTO, and we observe a close correspondence
between the 125\,cm$^{-1}$ modes of CdCTO and CCTO.  The latter two
are dominated by Cu-O motion, and thus are unaffected by the
replacement of Ca with Cd.  We note here that peculiar changes in
oscillator strengths with temperature are observed to occur for
this mode in both materials,\cite{homes2} indirect confirmation of
their similar character.

The conspicuous reduction in frequency of the 
135\,cm$^{-1}$ mode in CCTO to 72\,cm$^{-1}$ 
in CdCTO originates from both inertial and chemical effects.
The masses of Ca and Cd are
112.4 and 40.1\,amu respectively; on this basis alone, the frequency
could decline at most by a ratio of $\sqrt{M_{\rm
Ca}/M_{\rm Cd}}$=0.597.  Since Ca/Cu motion
accounts for only 38\% of the eigenvector character, the inertial
shift must actually be considerably smaller, and thus cannot account
for the calculated frequencies.  Indeed, we carried out a test
calculation in which the Ca mass was replaced by that of Cd while
retaining the force-constant matrix of CCTO.
Rediagonalizing the dynamical matrix, we observe that the
lowest four frequencies change to 102, 126, 155, and
186\,cm$^{-1}$, respectively (the other seven remain almost
unchanged).  Thus, the mass effect alone would only be able to
explain a decrease from 135\,cm$^{-1}$ to 102\,cm$^{-1}$, i.e., 
a downward shift of about 24\%.  This is much less than the reduction observed
experimentally (48\%) or theoretically (47\%).  The remainder of the
shift is associated with an appreciable softening ($\sim$~40\%) of
the force constants of this mode.  Evidently this softening is
connected with chemical differences between the two atoms, e.g., the
$4d$ shell present in Cd but absent in Ca.

%---------------------------------------------------------------------
\begin{table}
\caption{Comparison of calculated Born effective charge 
tensors $Z^*$ (in a Cartesian basis) for CCTO and CdCTO.
Atoms are at Wyckoff positions given in the caption of Table I.}
\label{tab:zstar}
\vskip 0.1cm
\begin{tabular}{ddddddd}
 & \multicolumn{3}{c}{CaCu$_3$Ti$_4$O$_{12}$} &
\multicolumn{3}{c}{CdCu$_3$Ti$_4$O$_{12}$} \\
\hline
Ca/Cd       & 2.46  & 0     & 0    & 2.47 & 0    & 0\\
            & 0     & 2.46  & 0    & 0    & 2.47 & 0\\
            & 0     & 0     & 2.46 & 0    & 0    & 2.47\\\hline
Cu          & 2.06  & 0     & 0    & 2.09 & 0    & 0\\
            & 0     & 1.85  & 0    & 0    & 1.87 & 0\\
            & 0     & 0     & 1.21 & 0    & 0    & 1.10\\\hline
Ti          & 6.98  & $-$0.13 & 0.06 & 6.91 &$-$0.18 & 0.03\\
            & 0.06  & 6.98  &$-$0.13 & 0.03 & 6.91 & $-$0.18\\
            & $-$0.13 & 0.06  & 6.98 & $-$0.18& 0.03 & 6.91\\\hline
O           & $-$1.92 & 0.54  & 0    & $-$1.94& 0.54 & 0\\
            & 0.22  & $-$1.94 & 0    & 0.25 &$-$1.91 & 0\\
            & 0     & 0     &$-$5.01 & 0    & 0    & $-$4.94\\
\end{tabular}
\end{table}
%---------------------------------------------------------------------

In our previous work,\cite{lixin} we calculated the mode effective charges
$Z^{*}_{\lambda}=V \Delta P_z/u_0$ for each $z$-polarized IR-active
mode of CCTO using the Berry-phase formalism.\cite{ksv} (Here
$\Delta P_z = P_z({\bf u}_{eq} - u_0 {\bf u}_{\lambda}) - P_z({\bf
u}_{eq})$, ${\bf u}_{eq}$ is the equilibrium configuration of ions,
and ${\bf u}_{\lambda}$ is the eigenvector of mode $\lambda$ and
$u_0$ is its amplitude.) We repeat the process for CdCTO and obtain
the mode effective charges; the corresponding oscillator strengths
are compared with experiment in Table II.  As we found for
CCTO, the individual oscillator strengths disagree with experimental
values for some of the modes.  Remarkably, however,
the mode that was absent from the CCTO spectrum reappears in the case
of the CdCTO spectrum (Ref.~\onlinecite{homes2}), albeit with
an oscillator strength of only 0.3;  for both materials
this mode is primarily composed of O ($\sim$~80\%) and Ti
($\sim$~15\%) displacements.

From the mode effective charges we now estimate the lattice
contribution to the static lattice
dielectric constant by summing over our calculated oscillator
strengths, i.e., $\epsilon_{\rm ph} = \Omega_0^2  \sum_\lambda
Z_{\lambda}^{*2} / \omega_{\lambda}^2 =\sum_\lambda S_\lambda$.
Using the eigenmodes and their effective charges in Table II, we
obtain $\epsilon_{\rm ph}$~=57.6, in
excellent agreement with the experimental value at room temperature 
($\epsilon_{\rm ph}$~=62.1) but somewhat less that the 
low temperature value ($\epsilon_{\rm ph}$~=106 at 10 K),
where an anomalous increase in the oscillator strength 
has been observed.\cite{homes} Compared with value of $\sim$\,45 computed for CCTO,
that obtained for CdCTO is $\sim$50\% larger, a consequence of
the lowest-frequency IR phonon in CdCTO having a lower frequency
and larger effective charge than the corresponding mode in CCTO.
Thus, despite differences in computed oscillator
strengths for these modes, their combined contribution to the
dielectric constant is remarkably close to that observed
experimentally. Since the computed frequencies agree much better
with experiment than the oscillator strengths, the 
agreement for $\epsilon_{\rm ph}$ implies some degree of error
cancellation in the latter.

To complete our comparative study of these two materials, we also
compute their atomic effective charge tensors.  This is done in
practice by starting with the mode effective charges and then using the
corresponding mode eigenvectors to transform back to an atomic
displacement basis.  (The results for CCTO were not reported in
Ref.~\onlinecite{lixin} and are given for the first time here.)
As can be seen from Table III,
the atomic effective charge tensors are virtually identical for both
materials,\cite{charge} and thus the same discussion applies to both.
The Ca and Cd cations have diagonal and isotropic effective charge
tensors of almost identical magnitude, +2.46 and +2.47
respectively. Thus, while the force constants are quite sensitive to the
different underlying atomic shell structure of the Ca and Cd ions,
the effective charge is not, suggesting that
there is little dynamical charge transfer between this cation and its
oxygen neighbors. For Cu, the two components associated with displacements
parallel to the CuO$_4$ plaquettes are close to the nominal valence
(+2), while the third component corresponding to normal displacements
is considerably smaller.
Except for the presence of some small off-diagonal components permitted
by symmetry, the results for Ti and O are strongly reminiscent of other
perovskites: we find that $Z^*$ for Ti has a large positive
anomalous component (relative to the nominal +4), and correspondingly
each oxygen has a negative anomalous component (relative to the nominal
$-2$) for displacements toward the nearest-neighbor Ti.  Indeed, the
atomic effective charges computed for Ca, Cd, Ti, and O are all
entirely consistent with those computed in other perovskite insulators
such as BaTiO$_3$\cite{zhong,ghosez} or CaTiO$_3$,\cite{cockayne} all
of which possess rather typical lattice dielectric constants of
$\epsilon_{\rm ph}$~$\sim$~10-100.

In summary, our first-principles calculations indicate that both
CCTO and CdCTO possess similar {\it intrinsic} structural,
vibrational, and dielectric properties.  Our computations of the
lattice contributions $\epsilon_{\rm ph}(0)$ to the static dielectric
constant are in good order-of-magnitude agreement with the values
measured experimentally in the far-infrared range for both CdCTO
and CCTO, but are drastically smaller than the enormous values
measured at frequencies below the Debye cutoff range.
The latter discrepancy reinforces the conclusion that some
extrinsic mechanism is likely to be responsible for the large dielectric
constant present in both materials.\cite{morrel}
Several important issues remain to be
resolved by experiment, such as the connection between internal sample
morphology and the large response, and the origin of the anomalous
behavior of the oscillator strengths of the Cu-O mode at low
frequency observed in both materials.

We would like to thank C. C. Homes and collaborators
for providing results prior to publication. We acknowledge
M. Marsman for his implementation of the Berry phase
technique within VASP. This work is supported by NSF Grant DMR-99-81193.

%%%%%%%%%%%%%%%%%%%%%%%%%%%%%%%%%%%%%%%%

%%%%%%%%%%%%%%%%%%%%%%%%%%%%%%%%%%%%%%%%


\begin{references}
\bibitem{volume} The lattice parameter of CCTO is obtained 
using a 37 Ry plane-wave cutoff, slightly larger than the value of
30 Ry used in Ref.~\onlinecite{lixin}, resulting in a slightly larger
CCTO lattice constant.  Better convergence is required here in
order to compare the nearly identical lattice parameters of CdCTO
and CCTO.
\bibitem{phonon} The slightly enlarged lattice constant
(Ref.~\onlinecite{volume}) gives rise to small downward shifts in
CCTO mode frequencies relative to Ref.~\onlinecite{lixin}, but they
are still in a excellent agreement with experiment; oscillator
strengths actually improve slightly with respect to experiment.
(GGA calculations at the GGA equilibrium lattice
constant result in phonon frequencies in considerably worse
agreement with experiment.)
\bibitem{charge} The average difference between the $Z^*$ tensors in
the two materials ($[ \sum_{\tau,ij}
\delta z_{\tau,ij}^2 / \sum_{\tau,ij}  \bar{z}_{\tau,ij}^2]^{-1/2}$)
is only 1.5\%.
\end{references}
\end{document}